# Evanescent absorption spectroscopy of transition metal dichalcogenide materials using guided photoluminescence in $Si_3N_4$ photonic waveguides

*Víctor Gonzalez-Morote[1*], Marianna Sledzinska[2], Arkadiusz P. Gertych[3], Syed Hassan Abbas Jaffery[3], Mariusz Zdrojek[3], Daniel Navarro-Urrios[1*]*

1 MIND-IN2UB, Department of Electronics and Biomedical Engineering, University of Barcelona, 08028 Barcelona, Spain

2 Catalan Institute of Nanoscience and Nanotechnology (ICN2), CSIC and BIST, Campus UAB, Bellaterra, 08193 Barcelona, Spain

3 Faculty of Physics, Warsaw University of Technology, Koszykowa 75, 00-662, Warsaw, Poland

On-chip integration of two-dimensional transition metal dichalcogenides (TMDs) with photonic waveguides underpins a growing class of compact optoelectronic and sensing devices, whose performance depends on the optical absorption of the 2D material in this integrated environment. We demonstrate a source-free, on-chip evanescent absorption spectroscopy of $MoS_2$ and $WS_2$ flakes integrated on $Si_3N_4$ waveguides, in which the broadband defect-related photoluminescence of the $Si_3N_4$ core itself acts as the internal probe. Light generated inside the core propagates under the TMD flake and is spectrally attenuated by evanescent interaction with the 2D material, and normalizing the transmitted spectrum to a reference waveguide without TMD yields the energy-dependent absorption response without any external illumination. The extracted guided absorption agrees with Raman and local micro-photoluminescence characterization and resolves differences between spatially uniform and non-uniform flakes. A complementary top-pumping scheme recovers guided $WS_2$ photoluminescence, whose spectral reshaping during propagation is quantitatively reproduced by the measured absorption. This source-free platform qualifies 2D materials in their integrated photonic environment and provides a direct route toward waveguide-integrated TMD photodetectors and sensors.

## 1. INTRODUCTION

Transition metal dichalcogenides (TMDs), such as $MoS_2$ and $WS_2$, are two-dimensional semiconductors which have attracted great interest because of their distinctive optical properties.[1] TMDs present strong thickness-dependent electronic band-gap transition, transitioning from an indirect to a direct bandgap in the monolayer limit, and exhibit pronounced excitonic resonances in the visible and near-infrared range.[2,3] Despite their atomic-scale thickness, monolayer TMDs display remarkably high optical absorption across the visible spectrum,[4] and their van der Waals nature enables their integration on arbitrary substrates without lattice matching requirements through dry-transfer techniques.[5] These properties make them attractive building blocks for compact and efficient optoelectronic devices integrated on photonic platforms.[6]

Integrating two-dimensional materials with photonic waveguides, from graphene photodetectors,[7] to transition-metal-dichalcogenide light emitters and detectors,[8] exploits the evanescent overlap between the guided mode and the atomically thin layer. By extending the interaction length along the propagation direction, the limited single-pass absorption is compensated, and the lateral absorption can approach unity over flake lengths of a few tens of micrometres.[9] Building on this principle, on-chip TMD functionalities on silicon nitride have enabled enhanced optical nonlinearity,[10,11] visible and near-infrared photodetection,[12,13] and mid-infrared all-optical modulation via interlayer-exciton absorption.[14]

In most of these works, the optical response of the 2D material is characterized either by free-space spectroscopy on the flake prior to integration,[15,16] or by monitoring changes in waveguide transmission using external broadband light sources coupled into the chip.[9] However, the optical response of a 2D material is not fixed. Strain, dielectric screening, and the flake-substrate interaction shift its excitonic transitions once the material is integrated.[17,18] As a result, characterization performed on the bare flake before integration does not necessarily reflect how the material behaves on the chip where it operates. This motivates a method that probes the absorption in situ, in the same waveguide and modal environment in which the material will function.

In this work, we introduce an alternative approach to probe the optical absorption of TMD layers ($MoS_2$ and $WS_2$) integrated on $Si_3N_4$ waveguides, based on on-chip guided photoluminescence (PL). Under above-bandgap illumination, the $Si_3N_4$ core itself acts as an internal broadband light source through defect-related PL, which couples into the guided modes and propagates along the waveguide. In the regions covered by the TMD flakes, the guided light is spectrally attenuated via evanescent interaction with the 2D material, providing a direct in-situ measurement of the guided-mode absorption spectrum without requiring external broadband illumination or complex coupling schemes. Because the probe is generated within the chip itself, the measurement scales naturally to many waveguides without additional broadband instrumentation, using only the pump laser that already excites the structure.

Beyond characterizing individual flakes, this method establishes a simple and scalable on-chip platform for qualifying 2D materials directly in the integrated photonic environment in which they operate and which ultimately governs device performance. When combined with microscopy, Raman, and μ-PL characterization of the individual flakes, it provides a tool for correlating material properties with the device-relevant optical response. The same absorption mechanism that underlies our measurement also constitutes a natural starting point for the future realization of integrated TMD-based photodetectors on $Si_3N_4$ platforms, which could be implemented using resistive or photoconductive device architectures.

## 2. EXPERIMENTAL SECTION

**2.1 $Si_3N_4$ Waveguide Platform.** The samples consist of amorphous $Si_3N_4$ ridge waveguides fabricated on a silicon substrate with a thermally grown SiO2 lower cladding. The $Si_3N_4$ core layer, deposited by low-pressure chemical vapor deposition (LPCVD), has a thickness of approximately 200 nm, while the waveguide widths range from 1 µm to 10 µm (see Figures 1a and 1b for a sketch of the samples). This reduced thickness restricts the vertical modal profile to the fundamental mode, such that all supported waveguide modes share a single antinode in the z-direction (perpendicular to the sample plane), while multimode behavior may still occur in the lateral direction. The evanescent tail of the optical modes extends into the air region above the waveguide surface and contains approximately 6-10% of the total modal energy (see Figure 1d), enabling efficient interaction between the guided light and the TMD flakes deposited on top of the structure. Due to the large width-to-height aspect ratio of the waveguides, this interaction is dominated by the upper waveguide interface, while the contribution of the lateral sidewalls to the overall modal overlap is negligible. The propagation loss of waveguides without TMD coverage is approximately 6 dB $cm^{-1}$ at a wavelength of 633 nm (see Supporting Information S4), significantly lower than the attenuation induced by the TMD flakes near their excitonic resonances.

**2.2 TMD Flake Transfer.** After fabrication, TMD flakes were transferred onto selected regions of the waveguides.

The detailed experimental procedure for the fabrication of 2D materials with gold-assisted exfoliation has been described in our previous work.[19] In brief, we started with flat bulk transition metal dichalcogenide (TMD) crystals (MoS2 and WS2) purchased from 2D Semiconductors (flux-grown crystals). A fresh crystal surface was exposed by removing the top layer using standard adhesive tape. Immediately after exfoliation, the freshly exposed crystal surface was brought into contact with a template-stripped flat gold layer prepared on a silicon wafer. To facilitate handling, the gold layer was coated with PMMA and attached to thermal-release tape. Separation of the gold layer from the bulk TMD crystal yielded large-area thin flakes, consisting predominantly of monolayers with occasional multilayer regions. The thin TMD layers were subsequently transferred onto substrates containing waveguides. The thermal-release tape was removed by placing the sample on a hot plate at 150 °C. PMMA was removed using acetone, and the gold layer was etched using a KI/I2 solution (Sigma-Aldrich standard gold etchant). The sample was then rinsed with deionized water. Finally, to further improve sample cleanliness, the samples were immersed in an acetone bath for 2 h, rinsed with isopropanol, and dried using a nitrogen gun.

The flakes only partially cover the waveguide surface, leaving several-hundred-micrometre-long sections of bare $Si_3N_4$ near the sample edges. These uncovered regions facilitate efficient optical coupling and allow broadband $Si_3N_4$ PL generated along the waveguide to accumulate in the guided modes before interacting with the TMD-covered sections.

**2.3 Guided-PL Absorption Spectroscopy.** Optical excitation is performed in a butt-coupling configuration using a lensed fiber (Figs. 1a, 1b). A continuous-wave laser at 405 nm (Cobolt 06-MLD, 3.06 eV photon energy), well above the effective bandgap of amorphous $Si_3N_4$, serves as the excitation source. Absorption within the waveguide core generates broadband PL attributed to radiative defect states characteristic of stoichiometric $Si_3N_4$.[20–22] A fraction of this broadband emission couples into the guided modes of the waveguide and propagates along the structure.

The light emerging at the output facet is collected by butt-coupling into a fiber, which is connected to a fiber collimator that launches the beam in free space into a Princeton Instruments Acton SP2750 monochromator coupled to a Princeton Instruments PIXIS:100 CCD camera (Fig. 1a). The 405 nm pump undergoes strong attenuation along the propagation path due to intrinsic absorption in $Si_3N_4$ and, more significantly, absorption in the TMD-covered sections, such that the detected signal in the relevant spectral window is dominated by the upstream-generated broadband $Si_3N_4$ PL. The guided PL interacts evanescently with the 2D material along the TMD-covered sections; normalization of the transmitted spectrum to that of a reference $Si_3N_4$ waveguide without TMD (Fig. 1c, left) isolates the spectral contribution of the integrated 2D material, yielding a direct measurement of its effective absorption response under guided-mode operation (Fig. 1c, right).

**2.4 Top-Pumping Excitation Configuration.** A complementary excitation scheme is also implemented in which the waveguide is top-pumped using a continuous-wave 491 nm laser (See Supporting Information S1). The beam is first expanded to a length of approximately 3 mm using a beam expander and subsequently focused into a narrow stripe along the propagation direction by a concave cylindrical mirror. The excitation stripe length is precisely controlled by partially clipping the beam with blades mounted on micrometric translation stages. This top-pumping configuration enables spatially selective excitation of arbitrary sections of the structure. When the excitation is localized near the output region of the waveguide, the PL generated in the last TMD-covered segments couples directly into the guided modes and reaches the output facet without undergoing sufficient reabsorption along the remaining propagation path to attenuate the signal below the noise floor. This scheme is used to record the guided $WS_2$ PL spectrum reported in Figure 3d.

**2.5 Micro-PL and Raman Characterization.** Local micro-photoluminescence (μ-PL) and Raman spectra were acquired in a backscattering configuration using a Horiba Jobin-Yvon LabRam HR 800 spectrometer coupled to an Olympus optical microscope, with a 532 nm diode-pumped solid-state laser focused through a ×100 objective. Spectra were recorded at multiple positions along each waveguide to map the spatial uniformity of the TMD flakes.

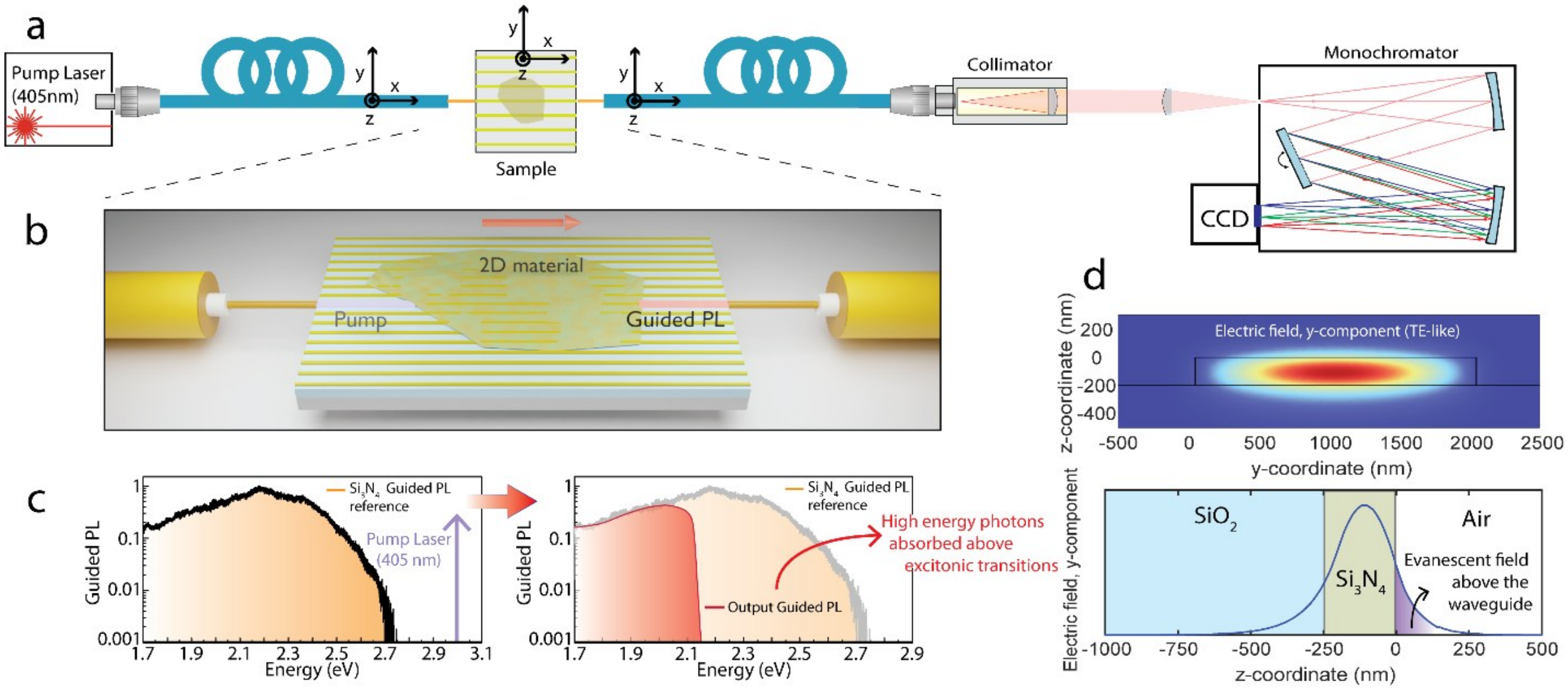

**Figure 1.** Evanescent absorption spectroscopy of a 2D material integrated on a dielectric waveguide. (a) Experimental configuration. A 405 nm pump laser is coupled into a $Si_3N_4$ waveguide and excites photoluminescence (PL) that is collected at the output via a fiber collimator and monochromator–CCD system. (b) Zoom of the experimental scheme. The broadband guided PL is spectrally filtered by evanescent absorption in the 2D-material-covered region. (c) Conceptual evolution of the guided PL spectrum. Left: reference spectrum corresponding to $Si_3N_4$ guided PL under 405 nm excitation. Right: modified spectrum after propagation under the 2D material, showing spectral reshaping due to evanescent absorption, particularly near the material's absorption edge. (d) Simulated optical mode profile. Top: transverse electric field distribution (fundamental TE-like mode) in the waveguide. Bottom: vertical refractive index profile and corresponding field distribution (almost equivalent for all supported modes), illustrating the evanescent tail overlapping with the 2D material region.

## 3. RESULTS AND DISCUSSION

**3.1 TMD Identification and Uniformity.** In this work we studied two representative TMD monolayers: $MoS_2$ and $WS_2$ (Figure 2b-c). For comparison, we also considered a few-layer $MoS_2$ flake (Figure 2a), with regions of thickness varying from one to three layers. Hereafter, these samples are referred to as uniform (monolayer) and non-uniform, respectively, according to their spatial thickness distribution.

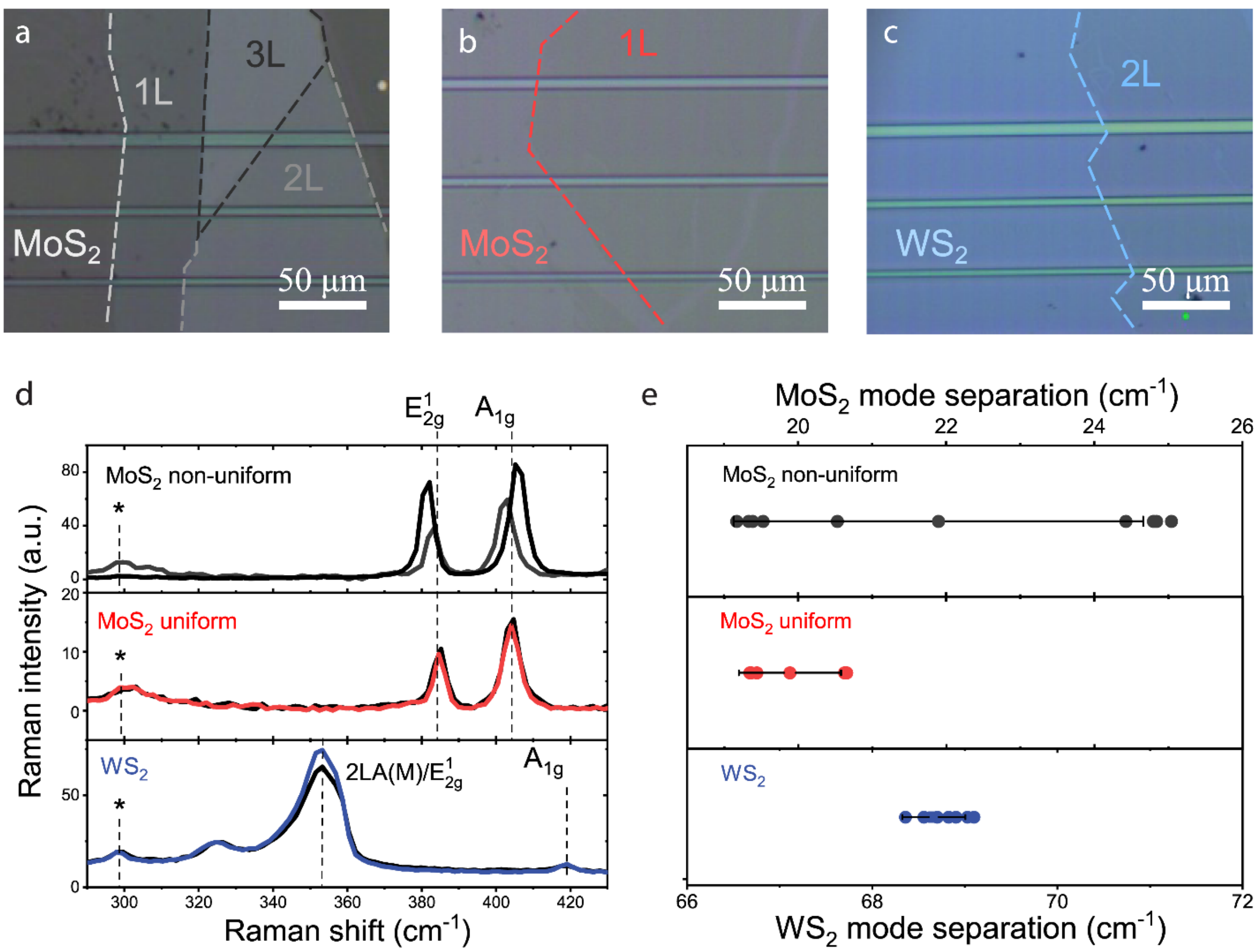

**Figure 2.** Optical and Raman characterization of the TMD-covered waveguides. (a–c) Optical microscopy images of the three studied samples: a spatially non-uniform $MoS_2$ flake containing regions with varying thickness, a uniform $MoS_2$ monolayer flake deposited on a series of waveguides; and a $WS_2$ monolayer flake. (d) Representative Raman spectra confirming the identity of the $MoS_2$ and $WS_2$ materials, respectively. The $MoS_2$ spectrum shows the characteristic $E_{2}g^{1}$ and $A_{1}g$ modes. The $WS_2$ spectrum displays the characteristic 2LA(M) / $E_{2}g^{1}$ band and the $A_{1}g$ mode. The asterisk marks two substrate-related Raman features around 300 and 520 cm-1. (e)

Raman peak separation Δω extracted from different positions on each sample. The individual data points and the corresponding mean and standard deviation provide a qualitative assessment of the local thickness variability and homogeneity of the samples.

Representative Raman spectra are shown in Figure 2d. The $MoS_2$ spectra show the in-plane $E_{2g}^1$ and out-of-plane $A_{1g}$ modes, while the $WS_2$ spectrum displays the convoluted 2LA(M)/$E_{2g}^1$ band together with the $A_{1g}$ mode, consistent with previous reports.[23,24]

The mode separation Δω plotted in Figure 2e corresponds to different physical quantities for the two materials. In $MoS_2$, $\Delta\omega = \omega(A_{1g}) - \omega(E_{2g}^1)$ is the well-established layer-number indicator, with values increasing from ~18 $cm^{-1}$ for monolayer to ~25 $cm^{-1}$ in the bulk limit.[24] For the uniform $MoS_2$ sample, the measured Δω clusters around 19.5 $cm^{-1}$ (Figure 2e), consistent with monolayer $MoS_2$. In $WS_2$, the well-established layer-number indicator is instead the intensity ratio $I(2LA(M))/I(A_{1g})$ under resonant excitation, since the 2LA(M) mode partially overlaps with $E_{2g}^1$ and the frequency separation between the two dominant spectral features is comparatively less sensitive to layer number.[24] The $WS_2$ Δω values shown in Figure 2e are therefore included only as a local homogeneity metric, and are not directly comparable to the $MoS_2$ Δω values, a distinction made explicit in Figure 2e by the use of separate abscissae for each material.

Within these constraints, the position-dependent Δω distributions reveal clear differences between the two $MoS_2$ samples: the non-uniform sample exhibits a broader distribution, consistent with stronger local thickness variability, whereas the uniform $MoS_2$ sample shows a narrower distribution. The $WS_2$ flake also displays a narrow Δω distribution, indicating spatial uniformity within the probed region.

**3.2 Guided-PL Extinction and Excitonic Absorption.** Figure 3 compares the spectrally filtered $Si_3N_4$ guided photoluminescence with local micro-photoluminescence (μ-PL) measurements, performed under a conventional top excitation–collection configuration. Since the broadband $Si_3N_4$ guided PL propagates underneath the 2D material, its spectrum becomes progressively attenuated by the optical absorption of all regions encountered along the propagation path. Consequently, the measured transmitted spectrum reflects the cumulative optical response of the material along the waveguide and is particularly sensitive to the lowest-energy optical transitions present in the structure.

The upper panel of Figure 3a corresponds to the non-uniform $MoS_2$ sample. The spatial heterogeneity of the TMD flake is directly reflected in the local μ-PL spectra acquired at different positions along the waveguide, which reveal significant variations in both emission energy and intensity. Regions associated with thicker material exhibit broader and red-shifted PL emission together with reduced intensity, consistent with the progressive evolution of $MoS_2$ optical properties with increasing thickness, including the transition from a direct-gap monolayer to an indirect-gap multilayer semiconductor.[2,3] The broad distribution of Raman peak separations observed in Figure 2e is therefore consistent with the pronounced spatial variations revealed by the μ-PL measurements.

In contrast, the guided $Si_3N_4$ photoluminescence experiences absorption by all regions of the flake encountered along the propagation path. As a result, the measured guided PL spectrum reflects the cumulative effect of a distribution of local optical transitions. In the non-uniform sample, the lowest-energy absorbing regions dominate the spectral filtering process, causing the guided-PL signal to be extinguished at 1.77 eV (black arrow in Figure 3a), well below the emission energies observed in other regions of the flake.

A markedly different behavior is observed for the uniform $MoS_2$ sample (central panel of Figure 3a). Consistent with the much narrower distribution of Raman peak separations observed in Figure 2e, nearly identical μ-PL spectra are obtained at different positions along the waveguide, indicating a spatially uniform material with relatively uniform optical properties. The spectra exhibit a single dominant emission band centered near 1.83 eV, commonly attributed to radiative recombination associated with the A exciton in monolayer $MoS_2$,[9,25] which is often the only peak visible in monolayer samples.[26] Accordingly, the guided-PL extinction energy appears at 1.81 eV (red arrow in Figure 3a), in close agreement with the local μ-PL emission energy.

Interestingly, this emission energy differs from that observed in the thinnest regions of the non-uniform $MoS_2$ sample. We attribute this shift to differences in the underlying waveguide geometry and local environment. In the non-uniform device, the $Si_3N_4$ waveguides underwent an additional $SiO_2$ etching step, resulting in a partially under-etched structure, whereas the uniform device was fabricated without a top $SiO_2$ layer. Such structural differences can modify the local strain distribution, dielectric screening, and flake–substrate interaction, all of which are known to influence the excitonic transition energies of atomically thin semiconductors.[26,27]

The lower panel of Figure 3a presents the corresponding measurements for $WS_2$ deposited on the same waveguide platform as the uniform $MoS_2$ sample. Consistent with the uniform optical contrast and the narrow distribution of Raman peak separations observed in Figure 2c and Figure 2e, the μ-PL spectra are highly reproducible at different positions along the waveguide, confirming the homogeneous optical properties of the flake. In this case, the guided-PL extinction energy is clearly shifted towards higher energies, occurring at 1.95 eV (blue arrow in Figure 3a), in agreement with the larger bandgap of monolayer $WS_2$ compared with $MoS_2$.

The μ-PL spectra display two partially overlapping emission bands. Based on previous reports on monolayer $WS_2$, these bands are explained in terms of the coexistence of neutral exciton (X) and negatively charged trion ($X^-$) recombination channels at room temperature.[28,29] The photon-flux dependence of both peaks and a more detailed analysis are provided in Supporting Information S2.

**3.3 Quantitative Attenuation Spectra and Detection Limits.** Figures 3b and 3c show the attenuation spectra extracted from the guided $Si_3N_4$ PL measurements for the uniform $MoS_2$ and $WS_2$ samples, respectively. The spectra are obtained by comparing the transmitted guided PL with the low-energy spectral region, where absorption by the TMD is assumed to be negligible. To express the attenuation in units of dB $cm^{-1}$, the measured absorbance was normalized by the effective length of the waveguide section covered by the TMD material. Coverage lengths of 680 μm for $MoS_2$ and 1.20 mm for $WS_2$ were extracted for the devices shown in Figure 3. It should be noted that these values correspond to the total extent of the TMD-covered region along the waveguide, as determined from optical microscopy images, and may include small uncovered gaps or discontinuities within the flake. As a result, the actual optical interaction length may be slightly shorter than the estimated coverage length, leading to a modest underestimation of the attenuation coefficient. The procedure used to determine the coverage length from optical microscope images is described in Supporting Information S3. The attenuation spectra shown are representative of all waveguides measured for each sample, which consistently yielded the same extinction features.

For the $MoS_2$ sample, remnants of the A and B excitonic resonances can still be identified as broad features centered near 1.84 eV and 2.01 eV, respectively, in agreement with previous reports.[26] In particular, the energy of the lower-energy feature coincides with the A-exciton emission observed in the local μ-PL measurements of the same flake (see central panel of Figure 3a), providing further evidence that the attenuation spectrum is governed by excitonic absorption in the TMD. In the $WS_2$ sample, the excitonic structure is less clearly resolved, although the onset and subsequent increase of the attenuation around 1.95–2.0 eV are consistent with the expected spectral position of the excitons. However, near the excitonic resonances, the attenuation rapidly approaches the dynamic-range limit of the experiment. As a consequence, the resonances are only partially resolved and their intrinsic spectral profiles cannot be fully reconstructed from the measured attenuation spectra. Beyond confirming the excitonic origin of the attenuation, the extracted spectra fix the two parameters that govern a waveguide-integrated photodetector.[9] The spectral band over which each material absorbs guided light defines the device operating range, and the absorption per unit length sets the length needed for efficient absorption.

At the guided-PL extinction energy, the transmitted guided PL approaches the experimental noise floor. Consequently, the maximum measurable attenuation is limited by the dynamic range of the experiment, defined as the ratio between the guided-PL signal and the noise floor. In the present measurements, this dynamic range is approximately 30 dB. For the TMD coverage lengths considered here, this corresponds to maximum measurable attenuation values of approximately 450 dB $cm^{-1}$ for $MoS_2$ and 250 dB $cm^{-1}$ for $WS_2$. The lower attenuation limit obtained for $WS_2$ results from its longer interaction length, which allows larger total optical losses to accumulate before the transmitted signal reaches the noise floor.

Most of the attenuation spectrum remains below these limits and can therefore be directly quantified. However, in the vicinity of the excitonic resonances, where absorption is expected to be strongest, the transmitted guided PL approaches the detection limit and the measured attenuation begins to saturate. As a result, the attenuation values reported at the excitonic resonances should be regarded as conservative lower-bound estimates, and the actual optical losses in these spectral regions may be significantly larger than those extracted from the measurements.

In comparable $Si_3N_4$ platforms, monolayer $MoS_2$ was shown to absorb guided light with losses close to 103 dB $cm^{-1}$ [9,25] near the A-exciton resonance. These values exceed the dynamic-range limit of the present measurements by a factor of 2-3, indicating that the true attenuation near the excitonic resonances is likely substantially larger than the values extracted here.

The comparatively low attenuation values extracted in the present work might, at first sight, be attributed to the simultaneous contribution of TE- and TM-like guided modes. Since the optical probe employed here is the intrinsic $Si_3N_4$ photoluminescence, which is not expected to exhibit a preferred polarization, both mode families could in principle contribute to the measured signal. This possibility is particularly relevant because excitonic absorption in TMDs is generally associated with in-plane electric-field components, leading previous studies to report substantially stronger attenuation for TE-like modes than for TM-like modes in tightly confined waveguide geometries.[30,31] To assess the role of polarization in our platform, the guided photoluminescence emitted from the waveguide facet was collected in free space using a microscope objective. A linear polarizer was then introduced in the collection path, immediately before the monochromator and CCD detector, to separately analyse the two orthogonal linear polarization components of the emitted signal. The attenuation spectra extracted from measurements performed without the polarizer and with each of the two polarization settings were found to be nearly identical, indicating that the absorption losses measured in the present devices are essentially independent of the detected polarization state of the guided photoluminescence. The corresponding measurements are presented in Supporting Information S6. Therefore, the lower attenuation values reported here cannot be explained by a simple averaging of strongly different TE- and TM-mode losses and are likely related to other factors. In particular, the finite dynamic range of the experiment causes the attenuation spectra to saturate as the transmitted guided-PL signal approaches the detection noise floor, so that the values extracted near the excitonic resonances should be regarded as conservative lower-bound estimates. A smaller additional contribution may arise from uncertainty in the effective interaction length due to small uncovered regions within the TMD-covered section.

In addition to the different absorption onsets, several lower-energy resonant-like features are also observed. These features appear at energies below the exciton emission energies identified in the μ-PL spectra, indicating that they cannot be directly assigned to the dominant excitonic recombination channels. Their origin is therefore likely associated with other absorbing states arising from localized regions of the flake experiencing different local environments, including variations in strain.[32] Such effects may be particularly pronounced near the lateral edges of the waveguide, where the mechanical and dielectric boundary conditions differ from those in the central region.

**3.4 Guided $WS_2$ Photoluminescence and Reabsorption.** The extracted attenuation spectrum provides a quantitative prediction for the spectral reshaping of guided $WS_2$ emission, which we test using a complementary top-pumping configuration. Figure 3d compares, for the $WS_2$ sample, four quantities: the extracted absorbance spectrum together with its polynomial fit, the local μ-PL spectrum, the guided $WS_2$ PL collected under top-pumping excitation, and the guided PL after correction for propagation reabsorption. The top-pumping configuration excites the $WS_2$ near the output facet of the waveguide, minimizing reabsorption of the emitted light during the remaining propagation length. Under equivalent excitation and collection conditions, $Si_3N_4$ waveguides without $WS_2$ exhibit guided PL intensities more than one order of magnitude lower, confirming that the measured guided emission is dominated by the $WS_2$ optical response.

The guided PL spectrum is systematically red-shifted with respect to the local μ-PL, reflecting the energy-dependent absorption landscape through which the $WS_2$ emission propagates before reaching the output facet. This reshaping is quantified by applying an inverse reabsorption correction to the guided-PL spectrum according to $I_{corr}(E)=I_{guided}(E)\cdot 10^{A(E)/10}$, where A(E) is the absorbance in dB/cm extracted from the same waveguide. Because the transmitted guided signal approaches the experimental noise floor at high attenuation values, the absorbance becomes increasingly noisy in the vicinity of the excitonic region. To estimate the effect of propagation losses on the guided PL spectrum, a smooth polynomial approximation of the measured attenuation was therefore constructed (dashed curve in Figure 3c and Figure 3d). The polynomial fit is used solely to provide a continuous estimate of the propagation losses and does not represent a physical model of the excitonic absorption lineshape. Under these conditions, the corrected spectrum recovers a peak position and lineshape associated with the trion emission, in good agreement with the local μ-PL. This agreement supports the interpretation that the guided PL corresponds to the intrinsic $WS_2$ emission spectrally filtered by excitonic absorption during propagation along the waveguide.

The high-energy peak commonly associated with the neutral exciton is strongly suppressed in the guided PL relative to the μ-PL. This behavior is qualitatively consistent with the excitation-power dependence of the neutral exciton and trion emission in n-doped monolayer $WS_2$, where the trion contribution becomes increasingly dominant at higher excitation densities.[28,33] A μ-PL power series acquired on the same flake (Supporting Information S2) confirms this trend. While a direct quantitative comparison between the guided-PL and μ-PL excitation conditions is not possible because the measurements were performed using different excitation wavelengths (491 nm and 532 nm, respectively), the observed spectral shape suggests that the guided-PL signal is collected under conditions where trion emission dominates the radiative response. Furthermore, the neutral exciton emission occurs at higher energies, where propagation losses are stronger due to enhanced excitonic reabsorption, further reducing its visibility in the collected guided signal.

No measurable guided PL was detected from the $MoS_2$-covered waveguides under equivalent top-pumping conditions. This is consistent with the substantially lower PL quantum yield reported for monolayer $MoS_2$ as compared with monolayer $WS_2$,[3] which places the on-chip detection of $MoS_2$ guided emission below the dynamic range of the present configuration.

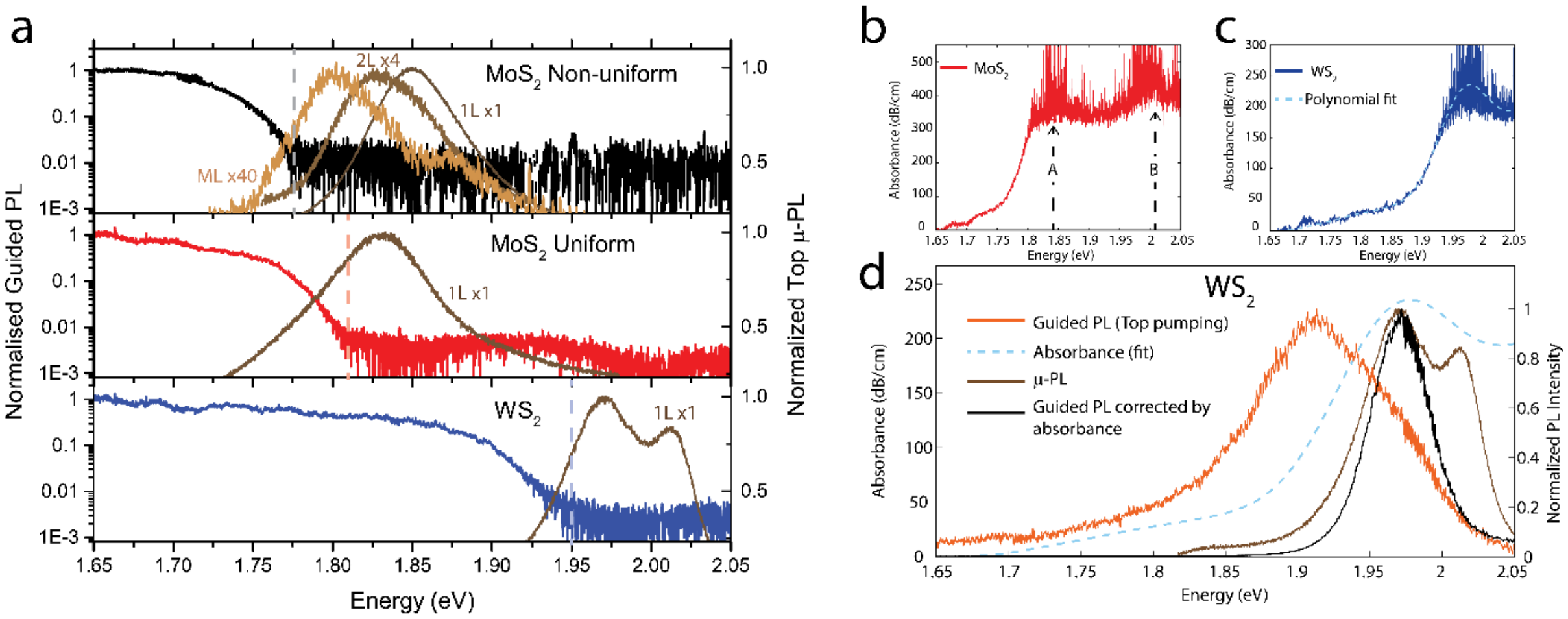

**Figure 3.** Evanescent absorption and guided photoluminescence of $Si_3N_4$ waveguides covered by TMD flakes ($MoS_2$ or $WS_2$). (a) Comparison between normalized $Si_3N_4$ guided photoluminescence transmitted through $MoS_2$ and $WS_2$ covered waveguides and local normalized μ-PL spectra acquired at different positions along the samples. The number of layers associated with each μ-PL spectrum and the corresponding normalization factor are also indicated (ML indicates an unknown number of layers). The upper and middle panels correspond to two different $MoS_2$ samples, while the lower panel corresponds to $WS_2$. Vertical dashed lines indicate the approximate position of the guided-PL signal extinction energy. (b) Attenuation spectrum (in dB $cm^{-1}$) extracted from the guided $Si_3N_4$ PL for the uniform $MoS_2$ sample. Arrows indicate the position of the excitonic resonances. (c) Same for the $WS_2$ sample, including a polynomial fit of the measured absorbance. In (b) and (c), the experimentally accessible attenuation range is limited to approximately 450 and 250 dB $cm^{-1}$, respectively. At larger attenuation values the transmitted guided PL becomes

indistinguishable from the detection noise floor, preventing a complete reconstruction of the intrinsic excitonic absorption profile. (d) Comparison, for the $WS_2$ sample, between the extracted attenuation spectrum, the local μ-PL spectrum, the guided-PL spectrum corrected using a fit of the extracted absorption losses, and the guided $WS_2$ PL collected through the waveguide under top excitation.

## 4. CONCLUSIONS

In summary, we have demonstrated an on-chip evanescent absorption spectroscopy technique for 2D semiconductors integrated on $Si_3N_4$ waveguides, in which the broadband defect-related photoluminescence of the $Si_3N_4$ core itself serves as an internal probe source. By normalizing the transmitted guided spectrum to that of a reference waveguide without TMD, the energy-dependent absorption response of $MoS_2$ and $WS_2$ flakes is extracted without requiring external broadband illumination or tunable light sources. The measurements reveal a clear correspondence between the guided-mode absorption response, Raman characterization, and local μ-PL spectra, enabling the distinction between spatially uniform and non-uniform flakes and providing direct information on the excitonic absorption properties experienced by guided modes. Furthermore, a complementary top-pumping configuration provides access to guided $WS_2$ photoluminescence, whose spectral reshaping during propagation is quantitatively explained by the independently measured absorption spectrum. The ability to recover the local μ-PL spectrum from the guided emission using the extracted attenuation profile provides direct validation of the proposed spectroscopy approach.

The measured absorption responses directly define the spectral window over which each 2D semiconductor can efficiently interact with guided light in this integrated platform, providing a practical input for the design of waveguide-coupled photodetectors and other on-chip optoelectronic devices based on TMDs. More broadly, the technique offers a simple route to relate local material characterization (Raman, μ-PL) to guided-mode operation under conditions directly relevant for integrated photonic devices.

Several natural extensions are envisaged. The principal limitation of the present implementation is the finite dynamic range imposed by the modest photoluminescence efficiency of stoichiometric $Si_3N_4$. Substituting the waveguide core with Si-rich silicon nitride would greatly enhance the available guided emission while keeping reasonably low losses, following the strategy successfully employed in active silicon nitride microdisks,[34] and thereby extend the accessible attenuation range of the technique. Furthermore, the same scheme can be applied to other 2D materials and to van der Waals heterostructures, providing a scalable diagnostic for the development of CMOS-compatible integrated photonic and optoelectronic devices based on 2D semiconductors.

## ASSOCIATED CONTENT

### Data Availability Statement

Data will be made available on request.

### Supporting Information.

The Supporting Information is available free of charge at https://pubs.acs.org/doi/XXXXX.

Top-pumping excitation configuration; power-dependent micro-photoluminescence of monolayer $WS_2$ and the exciton/trion analysis; determination of the TMD coverage length from optical microscopy images; propagation-loss characterization of bare $Si_3N_4$ waveguides; and polarization dependence of the guided-photoluminescence attenuation (PDF).

## AUTHOR INFORMATION

**Corresponding Authors**

**V. Gonzalez-Morote** –Department of Electronics and Biomedical Engineering, University of Barcelona, 08028 Barcelona, Spain; MIND Group, Institute of Nanoscience and Nanotechnology (IN2UB), University of Barcelona, 08028 Barcelona, Spain; orcid.org/0000-0001-6925-3135; Email: vgonzalezmorote@ub.edu

**Daniel Navarro-Urrios** − Department of Electronics and Biomedical Engineering, University of Barcelona, 08028 Barcelona, Spain; MIND Group, Institute of Nanoscience and Nanotechnology (IN2UB), University of Barcelona, 08028 Barcelona, Spain; orcid.org/0000-0001-9055-1583; Email: dnavarro@ub.edu

**Authors**

**M. Sledzinska** − Catalan Institute of Nanoscience and Nanotechnology (ICN2), CSIC and BIST, Campus UAB, Bellaterra, 08193 Barcelona, Spain; orcid.org/0000-0001-8592-1121

**A. P. Gertych** − Faculty of Physics, Warsaw University of Technology, Koszykowa 75, 00-662 Warsaw, Poland; orcid.org/0000-0002-7740-9651

**S. H. A. Jaffery** − Faculty of Physics, Warsaw University of Technology, Koszykowa 75, 00-662 Warsaw, Poland; orcid.org/0000-0003-2504-0124

**M. Zdrojek** − Faculty of Physics, Warsaw University of Technology, Koszykowa 75, 00-662 Warsaw, Poland; orcid.org/0000-0002-8897-6205

**Author Contributions**

V.G.-M. performed the optical measurements, carried out the data analysis, prepared the figures, and wrote the original manuscript. D.N.-U. conceived and supervised the study, contributed to the data analysis and interpretation, acquired funding, and reviewed and edited the manuscript. A.P.G.,

S.H.A.J. and M.Z. fabricated and transferred the 2D materials. M.S. contributed to the sample characterization and to the discussion of the results. All authors discussed the results and have given approval to the final version of the manuscript.

**Notes**

The authors declare no competing financial interest.

ACKNOWLEDGMENT

This work was supported by the MICIU through the projects ALLEGRO (Grant No. PID2021-124618NB-C22) and OMENS (Grant No. PID2024-156058NB-I00). ICN2 was supported by the Severo Ochoa Centres of Excellence Program, Grant CEX2021-001214-S, funded by MCIN/AEI/10.13039.501100011033. M.S. acknowledges support from the MICIU project ELEMENTAL (Grant No. PID2023-152783OB-I00), project PETITE (Grant No. PCI2023-143399) funded by MCIN/AEI/10.13039/501100011033 and by the European Union, and ICN2 Seed Fund Project STRIDE-MoD. APG and MZ thank for support of project OPUS NCN (2025/57/B/ST11/02260) and Warsaw University of Technology within the Excellence Initiative Research University programme, respectively

# Supplementary Information

## Evanescent absorption spectroscopy of transition metal dichalcogenides using guided photoluminescence in $Si_3N_4$ photonic waveguides

*V. Gonzalez-Morote[1], M. Sledzinska[2], A. P. Gertych[3], S.H.A. Jaffery[3], M. Zdrojek[3], D. Navarro-Urrios[1]*

[1] MIND-IN2UB, Departament d'Enginyeria Electrònica i Biomèdica, Facultat de Física, Universitat de Barcelona, Martí i Franquès 1, 08028 Barcelona, Spain

[2] Catalan Institute of Nanoscience and Nanotechnology (ICN2), CSIC and BIST, Campus UAB, Bellaterra, 08193 Barcelona, Spain

[3] Faculty of Physics, Warsaw University of Technology, Koszykowa 75, 00-662, Warsaw, Poland

## S1. Optical setup and excitation geometries

The complementary top-pumping scheme implemented in the waveguide characterization setup, briefly described in Section 2 of the main text, is sketched in Fig. S1. It uses a 491 nm continuous-wave laser (Cobolt Calypso), whose beam was first expanded by a factor of 5 using a confocal two-lens telescope with focal lengths of 25 mm and 125 mm. This produced a 3 mm collimated beam, whose diameter was verified using the knife-edge technique.[1,2] A concave cylindrical mirror (focal length 100 mm) then folded the beam onto the chip and focused it along one axis to a 10 µm line while leaving the orthogonal axis unfocused, thereby forming a 10 µm × 3 mm stripe oriented along the waveguide. The folded beam reaches the sample at an incidence angle of 55.5° with respect to the surface normal. Consequently, the illuminated area is increased by a factor of 1.77 relative to the nominal stripe area, and the photon-flux density at the sample surface is reduced by a factor of $\cos(\theta) = 0.57$. The alignment of the excitation stripe with respect to the target waveguide was monitored from above using a microscope objective and a CCD camera.

Guided photoluminescence (PL) generated by excitation of the TMD material deposited on the waveguides and emerging from the output facet was collected in a manner similar to that described for Fig. 1 of the main text.

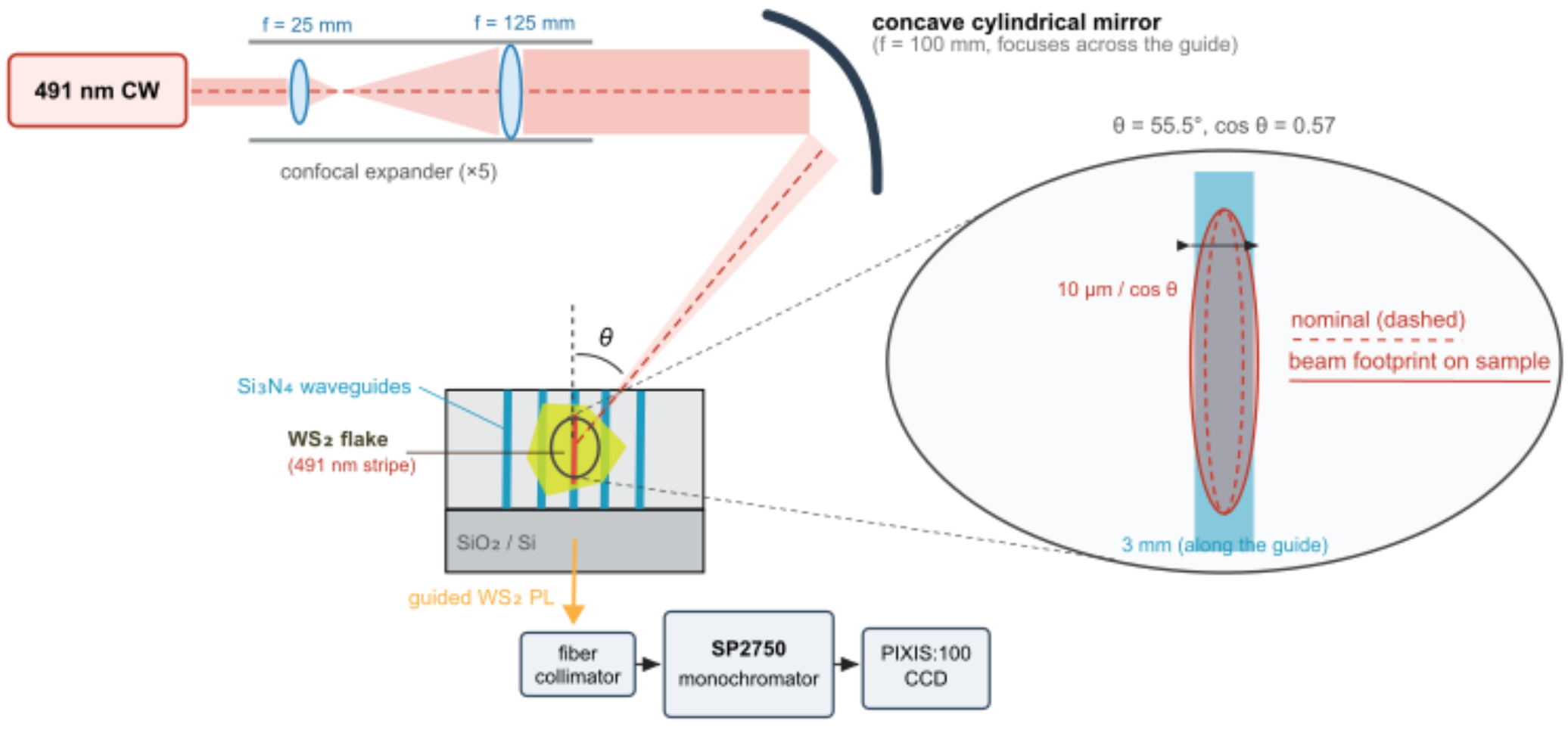

***Figure S1.*** *Top-pumping excitation geometry. A 491 nm continuous-wave laser is expanded ×5 by a confocal two-lens telescope (f = 25 mm and f = 125 mm) and folded onto the chip by a concave cylindrical mirror (f = 100 mm), which focuses it along one axis into a 10 µm × 3 mm stripe oriented along the waveguide. The beam reaches the sample at 55.5° from the surface normal, so its elliptical footprint (zoom) has the 10 µm width stretched to 10 µm/cos* $\vartheta$ *and the on-sample photon-flux density reduced by cos* $\vartheta = 0.57$ *(Sec. S4). Guided* $WS_2$ *PL from the output facet is collected by a fiber collimator and dispersed by an SP2750 monochromator onto a PIXIS:100 CCD.*

Conventional micro-photoluminescence (µ-PL) measurements were also performed using a top-excitation/top-collection geometry. In this configuration, a 532 nm continuous-wave laser was focused through a 100× microscope objective to a spot approximately 4 µm in diameter. A nominal laser power setting of 100% corresponds to 6 mW at the laser output and was adjusted using neutral-density filters. The same microscope objective was used to collect the µ-PL signal emitted by the TMD material, which was analyzed with a monochromator and a CCD.

All measurements were performed under ambient conditions.

**S2. Power-dependent photoluminescence, exciton and trion contributions to the guided emission of $WS_2$**

To better understand the guided $WS_2$ PL spectrum shown in Fig. 3d (orange curve), we compared two power-dependent PL series acquired on the same flake using the setups described in Section S1: (i) a conventional µ-PL series excited at 532 nm with powers ranging from 0.6 µW to 1.5 mW at the sample, and (ii) a guided top-pumping series excited with a 491 nm stripe beam with powers ranging from 12.5 mW to 125 mW at the sample.

In the µ-PL experiment, performed under normal-incidence excitation and therefore free from guided reabsorption effects, the neutral-exciton peak (X, 2.01 eV) dominates the emission spectrum at low excitation powers, whereas the negatively charged trion peak ($X^{-}$, 1.98 eV) becomes progressively more pronounced as the pump power increases (Fig. S2). Both contributions exhibit a sub-linear dependence on the incident photon flux $\Phi$ [photons $s^{-1}$ $cm^{-2}$], with effective power-law exponents of $\alpha_x = 0.68$ and $\alpha_{x-} = 0.75$, respectively (Fig. S3a).[3] Because the trion emission increases slightly faster than the neutral-exciton emission, the relative trion contribution grows with increasing excitation density and eventually becomes comparable to, or larger than, the exciton contribution at the highest accessible powers. This trend suggests that the trion dominance would be further enhanced at even higher excitation densities.

The integrated guided PL signal also follows a sub-linear dependence on photon flux, with a power-law exponent $\alpha_{guided} = 0.69$ (Fig. S3b), comparable to those obtained from the µ-PL measurements. Although the nominal photon-flux densities shown in Figs. S3a and S3b are of similar magnitude, the guided-PL spectrum presented in Fig. 3d exhibits a substantially stronger trion contribution. A likely explanation is the higher absorption efficiency of $WS_2$ at the 491 nm excitation wavelength (2.53 eV) compared with 532 nm (2.33 eV). The larger absorption cross section at higher photon energies results in a higher density of photogenerated carriers for a given incident photon flux, thereby favoring trion formation and reinforcing the trion-dominated emission observed in the guided-PL spectrum.

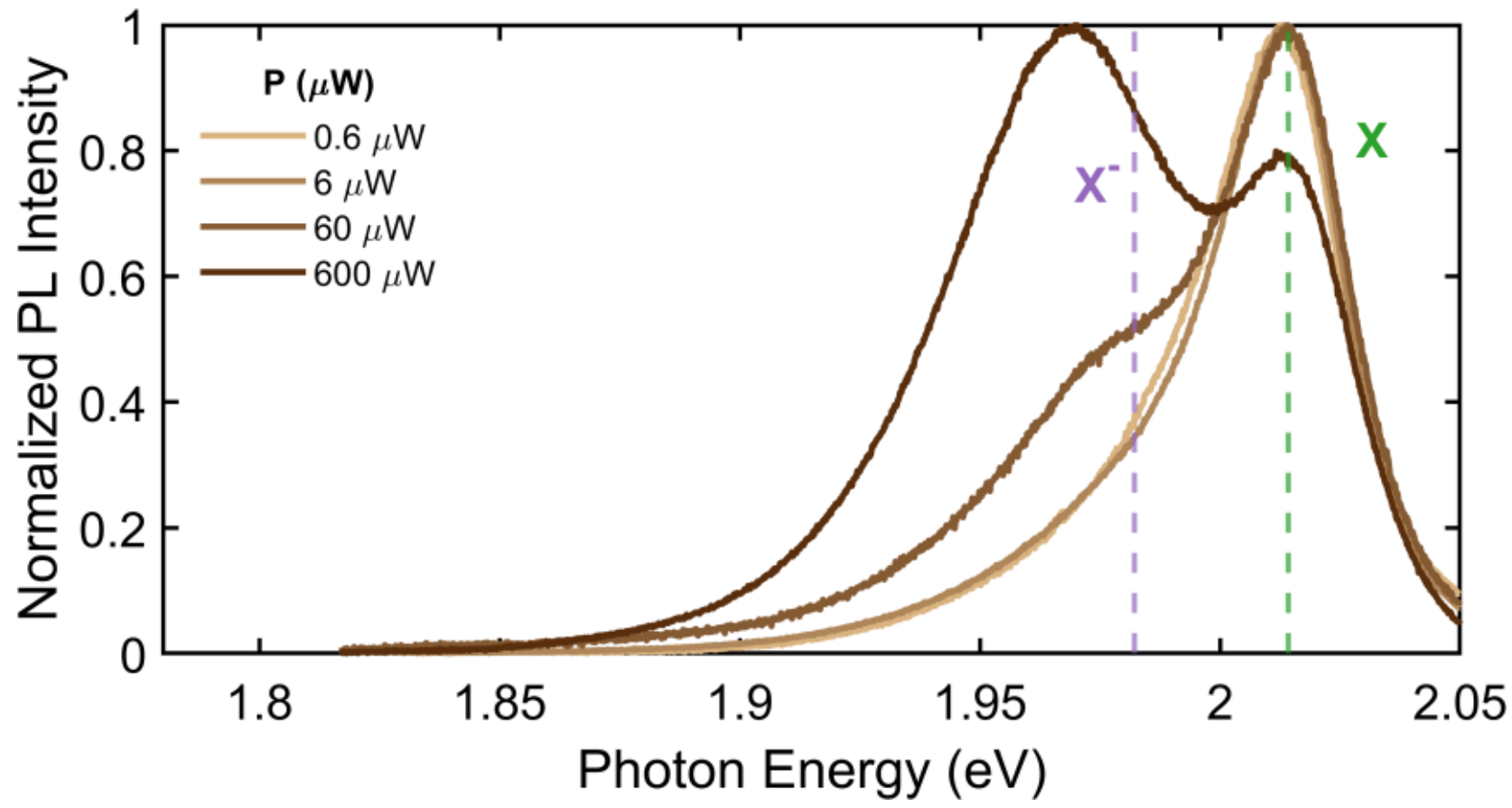

***Figure S2.*** *Normalized micro-photoluminescence (µPL) spectra of $WS_2$ (532 nm, ×100 objective) for excitation powers 0.6 µW–1.5 mW. The neutral exciton (X, 2.01 eV) and charged trion ($X^{-}$, 1.98 eV) are resolved; the trion grows faster with power and dominates at high flux.*

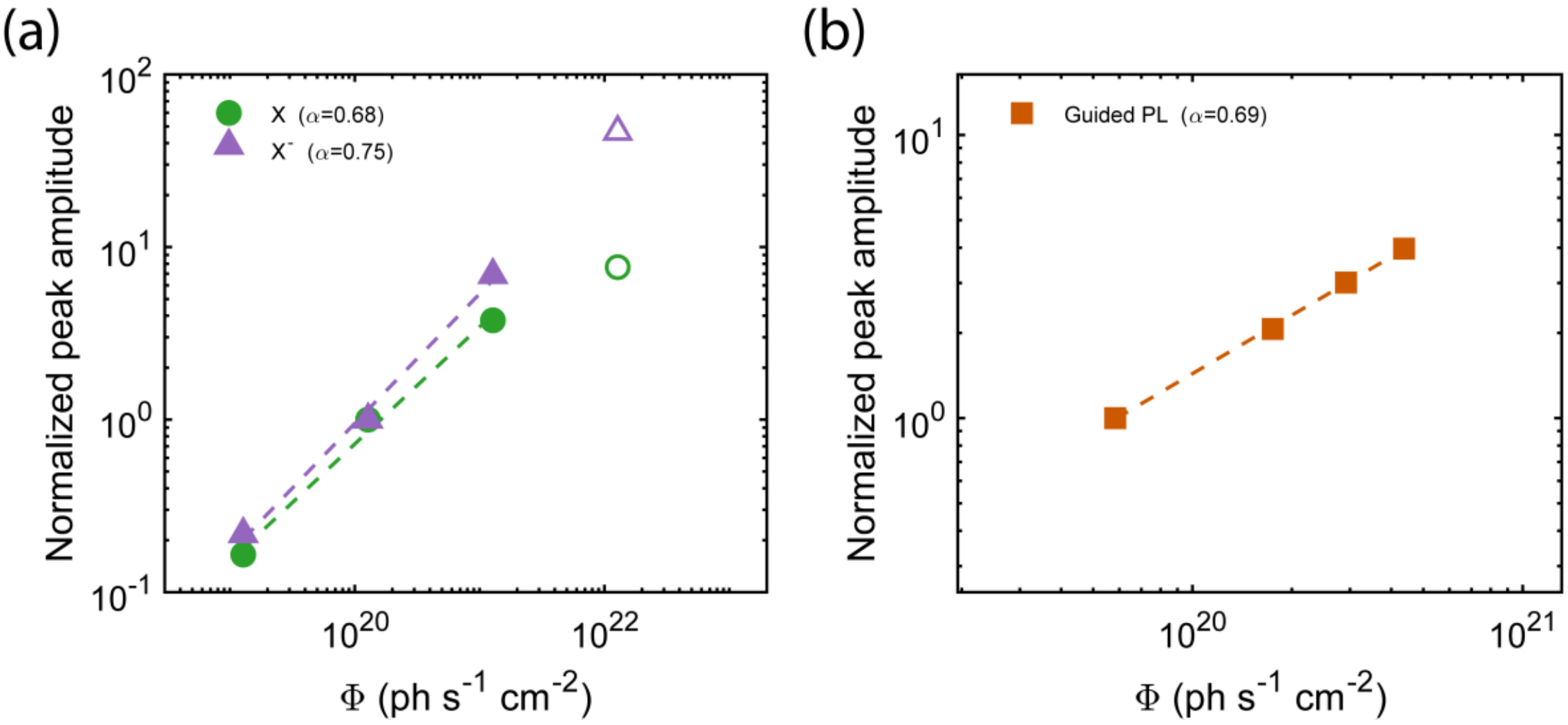

***Figure S3.*** *Normalized PL amplitude versus photon-flux density $\Phi$ (log–log). (a) The µ-PL exciton ($\alpha_X$ = 0.68) and trion ($\alpha_{X^-}$ = 0.75) follow distinct sub-linear power laws; (b) the guided PL ($\alpha_{guided}$ = 0.69).*

### S3. TMD coverage-length measurement

The attenuation spectra in the main text (Fig. 3b,c) are expressed per unit length, in dB cm$^{-1}$, obtained by normalizing the total absorbance accumulated over the TMD-covered section by the coverage length L. To measure L, a 405 nm beam was coupled from one facet and a 633 nm beam from the other, both strongly absorbed in the TMD-covered region. Each guided beam is therefore extinguished at the corresponding flake edge, and the dark gap between the two extinction points marks the covered length L (Fig. S4). Calibrating the image against the known 634 µm chip-block pitch converts this gap into an absolute length, giving L = 680 µm for the $MoS_2$ flake and L = 1.2 mm for the $WS_2$ flake.

The total absorbance A (in dB) yields an attenuation per unit length $\alpha\left(\frac{dB}{cm}\right)$.

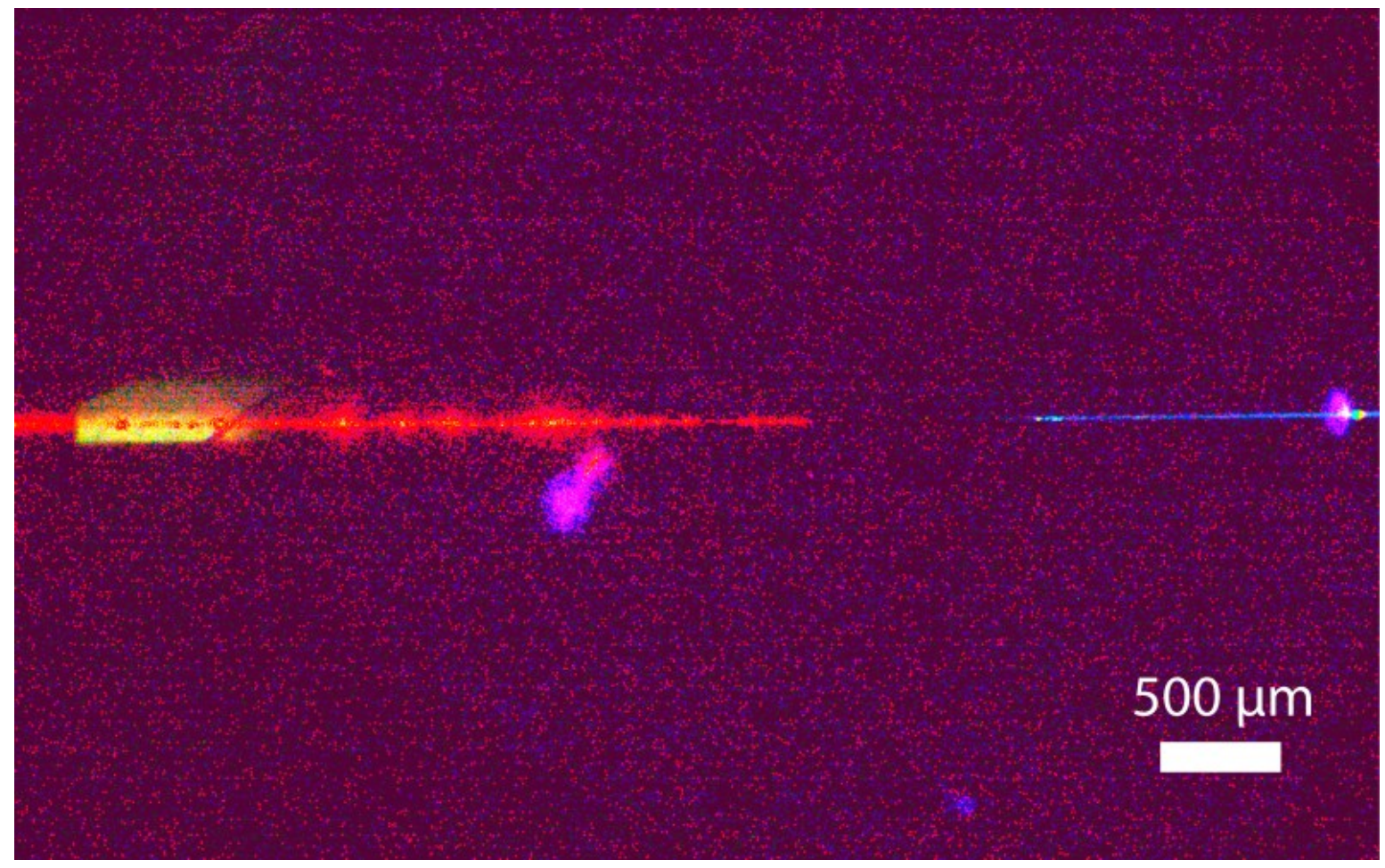

***Figure S4.*** *Coverage-length measurement of the uniform $MoS_2$ flake on a 10-µm-wide waveguide, with light coupled simultaneously from both facets (405 nm from one side, 633 nm from the other), recorded at high exposure. The two counter-propagating guided beams and the dark gap between their extinction points are clearly visible; this gap marks the TMD-covered length. Using the 634 µm chip block pitch as the in-frame scale gives L = 680 µm for the $MoS_2$ flake; the $WS_2$ flake was measured analogously, giving L = 1.2 mm.*

### S4. Waveguide losses

The scattered-light imaging technique is a non-destructive, top-view method commonly used to estimate propagation losses in optical waveguides and photonic integrated circuits. The technique relies on collecting the weak optical signal scattered out of the guiding plane as light propagates along the waveguide. Scattering originates from imperfections such as sidewall roughness, structural defects, and material inhomogeneities.

A high-resolution camera records the spatial distribution of the scattered light from above the chip (top panel of Fig. S5). The intensity profile along the propagation direction, z, is then extracted from the image (main panel of Fig. S5). Assuming that the scattered intensity is proportional to the guided optical power and that the scattering coefficient remains approximately constant along the waveguide, the propagation loss can be determined by fitting the measured intensity profile to an exponential decay:

$$I(z) = I_o e^{-\alpha z}$$

where $\alpha$ is the attenuation coefficient. The loss in decibels per centimeter (dB/cm) is then calculated as: $\alpha\left(\frac{dB}{cm}\right) = 4.34\alpha(cm^{-1})$. For the reference 10-µm-wide $Si_3N_4$ waveguide investigated here, the extracted propagation loss is about 6 dB/cm at 633 nm, corresponding to the spectral region where the TMD materials studied exhibit both absorption and photoluminescence. This value is more than one order of magnitude smaller than the attenuation measured in TMD-covered waveguides near the excitonic absorption edge, confirming that the optical losses reported in the main text are dominated by the interaction between the guided mode and the TMD layer rather than by intrinsic propagation losses of the $Si_3N_4$ platform.

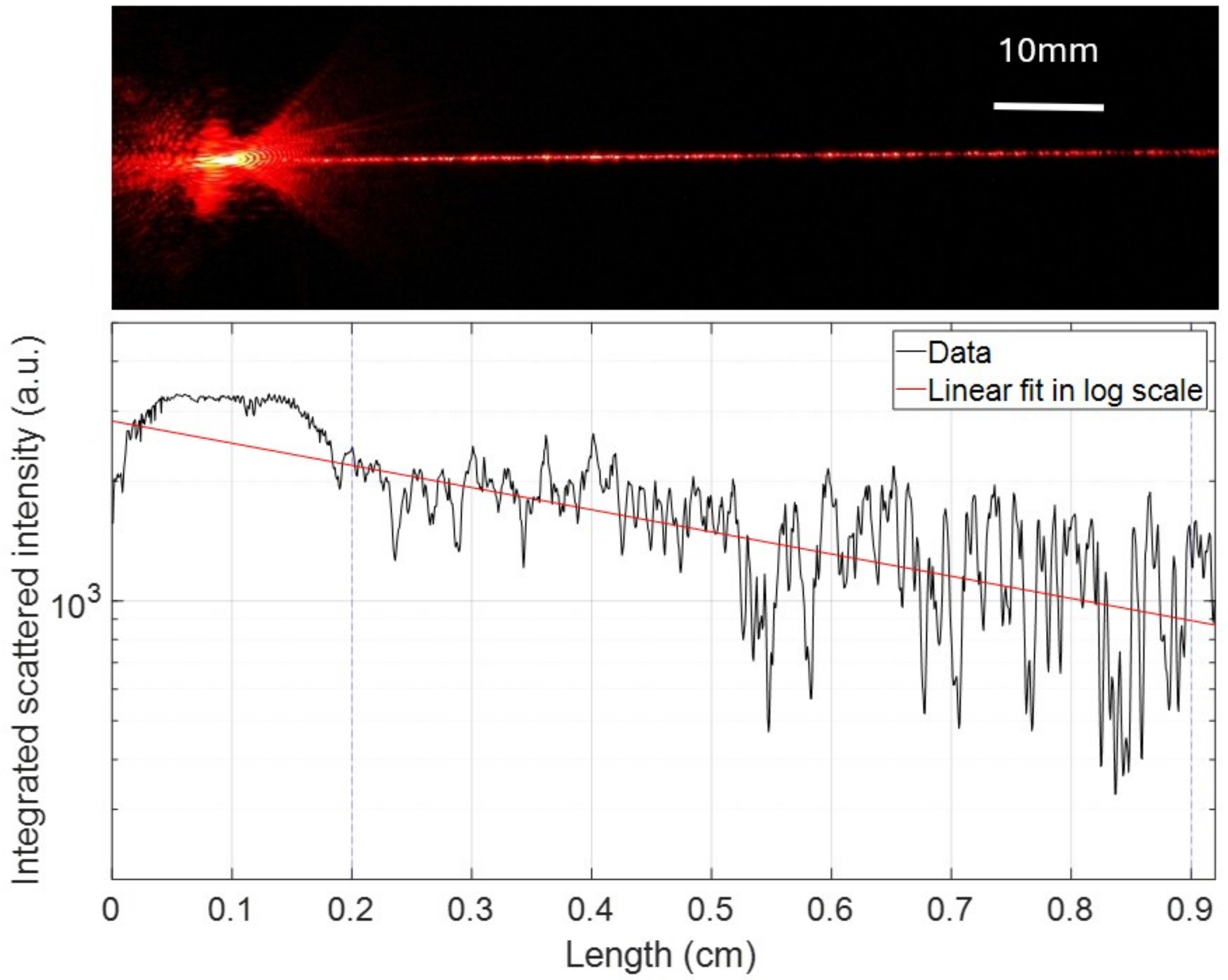

***Figure S5.*** *Propagation loss measurement of a 10-µm-wide reference $Si_3N_4$ waveguide without TMD material, obtained using the scattered-light imaging technique. The top panel shows the scattered-light image acquired from above the chip, while the bottom panel presents the extracted intensity profile and exponential fit used to estimate the propagation loss (about 6dB/cm at 633 nm).*

## S6. Polarization dependence

To investigate the possible influence of polarization on the extracted attenuation spectra, the guided photoluminescence emitted from the waveguide facet was collected in free space using a microscope objective. A linear polarizer was introduced in the collection path immediately before the monochromator and CCD detector to analyse two orthogonal linear polarization components of the emitted signal. Attenuation spectra were extracted from measurements performed without polarization selection and with the polarizer set to each of the two orthogonal orientations. As shown in Fig. S6, the three spectra exhibit very similar attenuation values and spectral evolution over the entire investigated energy range. These results indicate that, under the experimental conditions employed here, the extracted attenuation spectra are largely insensitive to the detected polarization state of the guided photoluminescence.

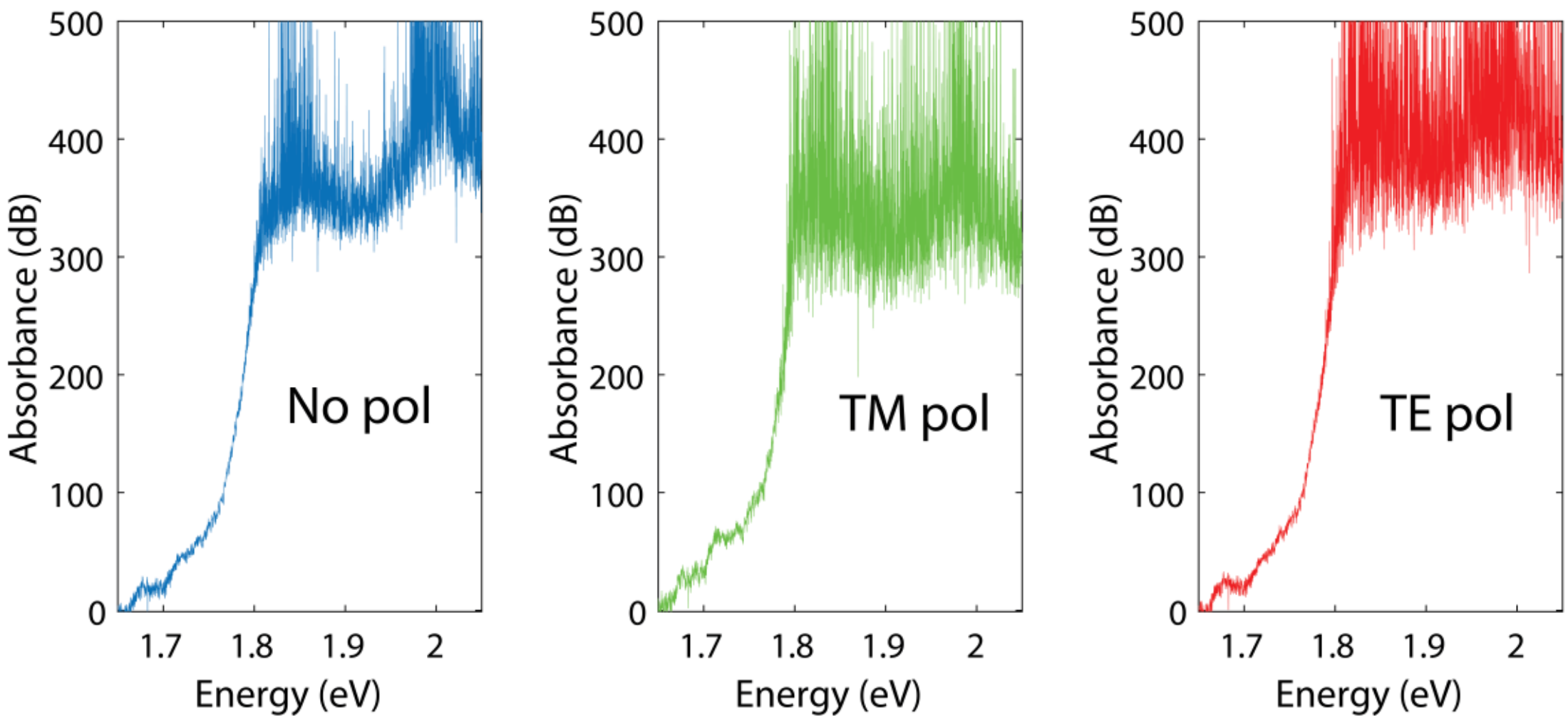

***Figure S6.*** *Attenuation spectra extracted from guided photoluminescence measurements acquired without polarization selection (left panel) and with two orthogonal linear polarization settings in the detection path (middle and right panels). The guided photoluminescence emitted from the waveguide facet was collected in free space using a microscope objective, while a linear polarizer placed immediately before the monochromator and CCD detector was used for polarization analysis. The three attenuation spectra exhibit comparable values and similar spectral evolution, indicating that the extracted absorption losses are essentially independent of the detected polarization state under the experimental conditions employed in this work.*